\documentclass{PoS}
\usepackage{amssymb,amsmath}
\usepackage{mwe}
\usepackage{subfig}
\usepackage[absolute]{textpos}

\title{Constraining QCD multijet background in the \textit{t}-channel single-top quark production at $\sqrt{s}=13\ TeV$}

\ShortTitle{In situ estimation of the QCD multijet background}

\author{\speaker{Georgios Konstantinos Krintiras} on behalf of the CMS Collaboration\\
        Universit\'e catholique de Louvain, Louvain-la-Neuve, Belgium\\
        E-mail: \email{gkrintir@cern.ch}}

\abstract{
Precision measurement of the cross section for single top production is an important test of the
Standard Model (SM). The purity of the collected data in single top events is limited by the understanding 
of the shape and yield of background contributions. Besides electroweak and $\rm{t\bar{t}}$ 
processes, QCD multijet events constitute a non-negligible background for the considered signal bq$'\rightarrow$ tq (\textit{t}-channel)
process. The data-driven technique for constraining QCD contribution, employed in the measurement 
of the \textit{t}-channel single top-quark cross section using the very first LHC proton-proton
collisions at $\sqrt{s}=13\ TeV$ with the CMS detector, is described. The dataset corresponds to an
integrated luminosity of $\mathcal{L}=42\ pb^{\mathrm{-1}}$.}

\FullConference{8th International Workshop on Top Quark Physics, TOP2015\\
		14-18 September, 2015\\
		Ischia, Italy}

\begin{document}

\section{\textit{t}-channel signature}

\begin{figure}[!ht]
  \centering
  \includegraphics[scale=0.35]{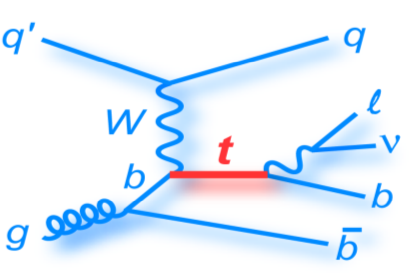}
    \caption{A representative diagram for the electroweak production of top quarks via \textit{t}-channel.}
    \label{fig:t_channel}
\end{figure}

The electroweak \textit{t}-channel production of top quarks (Fig. \ref{fig:t_channel}) is the most abundant single top
production mode in the Standard Model (SM) with a predicted cross section at approximate next-to-next-to-leading order (NNLO) 
that equals to $\sigma_{t\textrm{-ch.}}=214.5^{+1.0}_{-1.2}\ (\rm{scale})$$ pb$ at $\sqrt{s}=13\ TeV$ \cite{cite_key1}. More specifically, the signature of single top quark events 
produced via \textit{t}-channel and with a leptonic top-quark decay is dictated by the presence of exactly 
one \textbf{isolated lepton} (\textit{l}), a \textbf{\textit{b}-tagged jet} (\textit{b}), and \textbf{escaping neutrinos} ($\nu$) from the top quark decay chain, 
as well as a \textbf{light flavour jet} ($q$) recoiled mainly in the forward region. It should be noted that a second \textit{b} quark 
coming from gluon splitting results in an \textbf{additional \textit{b}-tagged jet} ($\bar{b}$) with a softer $p_T$ spectrum and a broader $\eta$ distribution.

Events are selected according to the above topology and divided in categories according to the number of jets and \textit{b}-tagged jets using
the naming convention of \textit{N}-jet-\textit{M}-tag, referring to events with \textit{N} jets, \textit{M} of which are \textit{b}-tagged \cite{cite_key2}.
The category enriched with \textit{t}-channel signal events is referred to as 2-jet-1-tag (2j1t). Exclusively
for this category, an additional requirement is applied on the mass of the reconstructed top quark
to lie within the $130<m_{l\nu b}<225\ GeV$ range. The observed and expected event yield, inside (SR)
and outside (SB) the $m_{l\nu b}$ window, is summarized below at the recorded $\mathcal{L}=42\ pb^{\mathrm{-1}}$, after imposing 
the $\rm{m_T^W}>50$ $GeV$ requirement on the transverse W-boson mass (Table \ref{tab:yield}). The systematic uncertainty
accounts for roughly $10\%$ of the final uncertainty on the estimated QCD yield. Possible sources of \textit{non-prompt} or misidentified leptons render 
QCD multijet events signal-like (cf. Sec. \ref{sec:Multijet_model}). 

 \begin{table}[h!]
 \centering
 \begin{tabular}{ |c|c||c| }
\hline
Process & SR & SB  \\
\hline
 $\rm{t\bar{t}}$ \& tW&$157\pm 1$&$71.7\pm 0.4$\\
W/Z+jets&$40\pm 4$&$47\pm 4$\\
QCD&$10\pm 5$&$2\pm 1$\\ 
\hline
$t$-channel&$33\pm 1$&$7.2\pm 0.3$\\
\hline
Total expected&$240\pm6 $&$128\pm 4$\\
\hline
Data & 252 & 127  \\
\hline
 \end{tabular}
\caption{\small{Event yields for the signal and the main background processes in the 2-jet-1-tag sample inside (SR)
and outside (SB) the $m_{l\nu b}$ window \cite{cite_key3}. Only top quark decays to muons are considered during the current study. 
Except for the number of QCD multijet events, all yields are taken from simulation and their uncertainties 
correspond to the size of the simulated samples. The yield of QCD multijet events is determined from the data.}}

\label{tab:yield}
 \end{table}

\section{Multijet model from data}
\label{sec:Multijet_model}
\begin{figure}[!ht]
  \centering
  \includegraphics[scale=0.2]{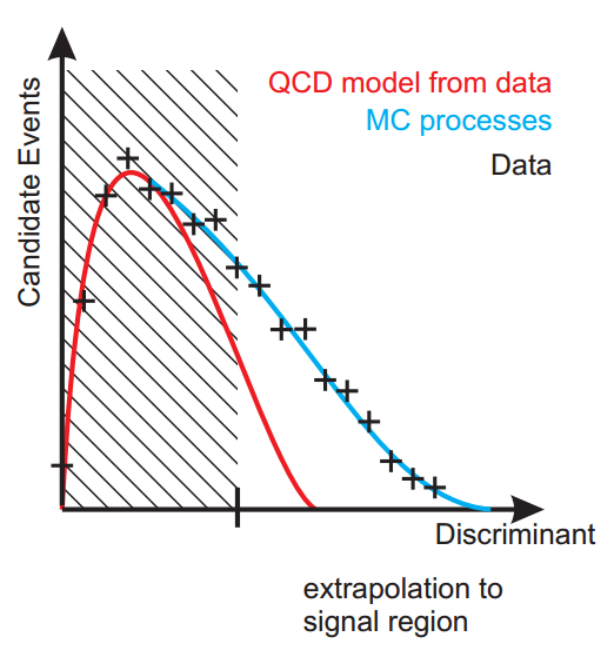}
    \caption{\small{Fictitious example that establishes the basic principle behind QCD estimation. The QCD shape
distribution is extracted by data over a phase space in the proximity of the signal region. The total amount of
QCD events is then determined by a template fit having utilized a variable that discriminates QCD against
non-QCD processes.}}
    \label{fig:principle}
\end{figure}

Given the associated theoretical uncertainties of QCD modelling, for instance stemming from
a not precise enough knowledge of the cross section or the approximate description of the underlying hadronic activity, 
it is necessary to predict the size and properties of this process by data.
A reliable model for determining QCD contamination is then derived by fitting templates of all physics processes to the
$x\ (\equiv \rm{m_T^W})$ discriminant (Fig. \ref{fig:principle}), i.e.
\begin{equation}
F(x)=N_{\rm{QCD}}\times Q(x)+N_{\rm{non-QCD}}\times W(x)\ .
\label{eq:fit}
\end{equation}

\hspace{-1.8em}The $W(x)$ templates, representing the signal and other non-QCD background processes, are taken
from the Monte Carlo simulation. After having subtracted non-QCD contamination, the $Q(x)$ template 
is constructed from a dedicated control sample that will be described in the following.

QCD multijet events become signal-like when an additional lepton is presented in the final state,
where a lepton could simply be the decay product of a hadron formed within jets. Typically, such
\textit{non-prompt} leptons, as opposed to leptons having been produced from the hard scattering, are
surrounded by many activities within the jet. On the other hand, leptons arisen out of the W-boson
leptonic decay are expected to be isolated, meaning that they should be found in a restricted spatial
region where they contribute with a major fraction in the deposited energy. Therefore a QCD
enriched data sample, yet orthogonal with the signal region, is achieved by referring to leptons
failing the isolation requirement.

Relative isolation, $I_{\rm{rel}}$, is defined in this analysis as 
\begin{equation}
I_{\rm{rel}}  = \frac{I^{\mathrm{ch.\,h}}+max((I^{\gamma}+I^{\mathrm{n.\,h}}-I^{\mathrm{PU}}),0)}{p_T}\ ,
\end{equation}

\hspace{-1.8em}where $I^{\mathrm{ch.\,h}}$, $I^{\gamma}$ and $I^{\mathrm{n.\,h}}$ are the sum of the transverse energies of stable particles in a cone $\Delta R = 0.4$
around the lepton direction with momentum $p_T$; $I^{\mathrm{PU}}\equiv \Delta \beta \times \sum p_T^{\mathrm{PU}}=0.5\times \sum p_T^{\mathrm{PU}}$
is the sum of transverse momenta of tracks associated to pileup vertices. The $\Delta \beta$ prefactor corresponds to
the naive expectation about the average of neutral to charge particle ratio and corrects the total
contribution of neutral $I^{\gamma}+I^{\mathrm{n.\,h}}$ candidates for pileup contamination. Such correction is validated
by monitoring the stable behavior of $I^{\gamma}+I^{\mathrm{n.\,h}}$ against the number of overlapping collisions.

\section{Results}
 \begin{figure}
\begin{minipage}{.5\linewidth}
\centering
\subfloat[]{\label{main:a}\includegraphics[width=6cm,height=4.5cm]{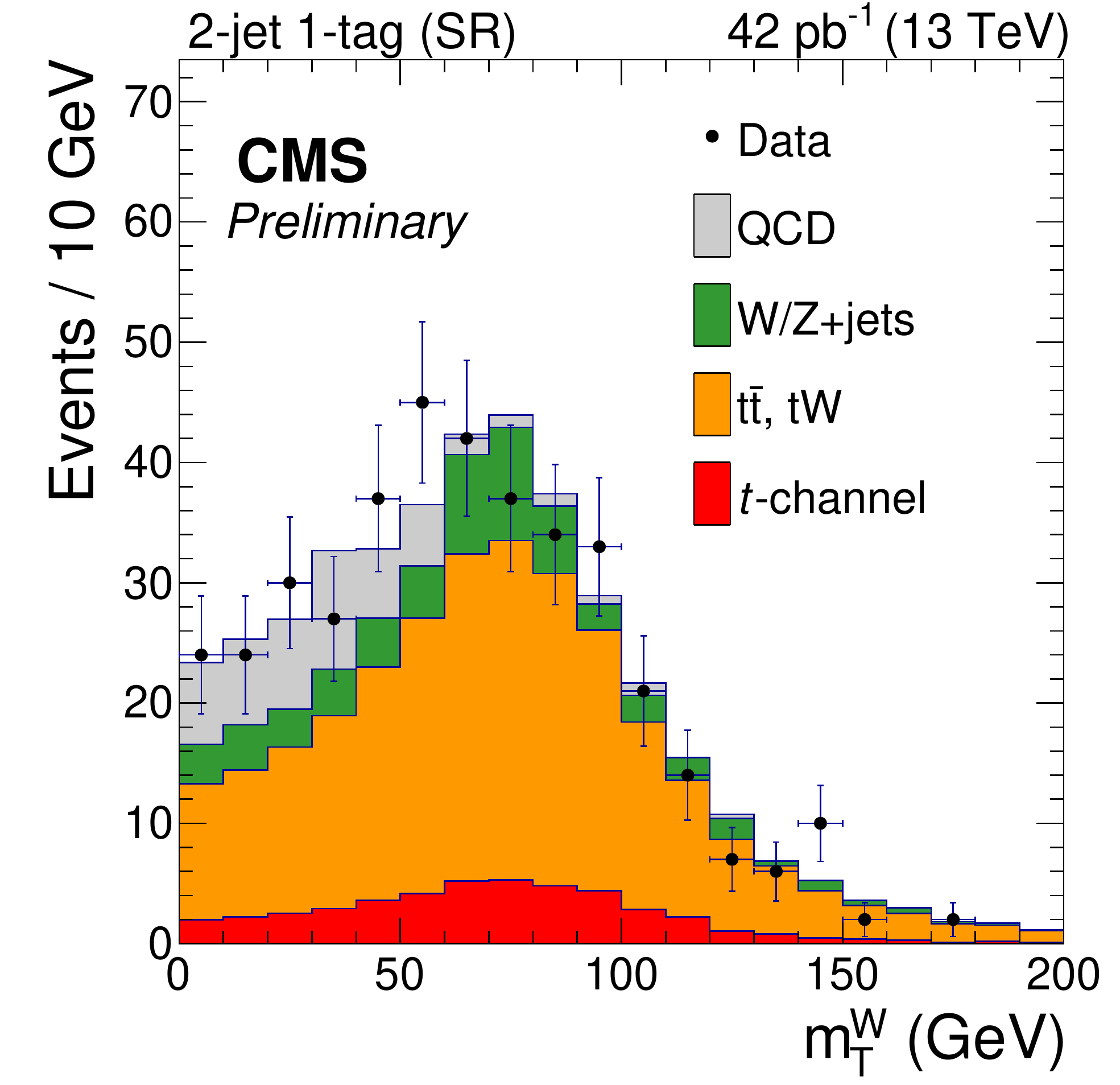}}
\end{minipage}%
\begin{minipage}{.5\linewidth}
\centering
\subfloat[]{\label{main:b}\includegraphics[width=6cm,height=4.5cm]{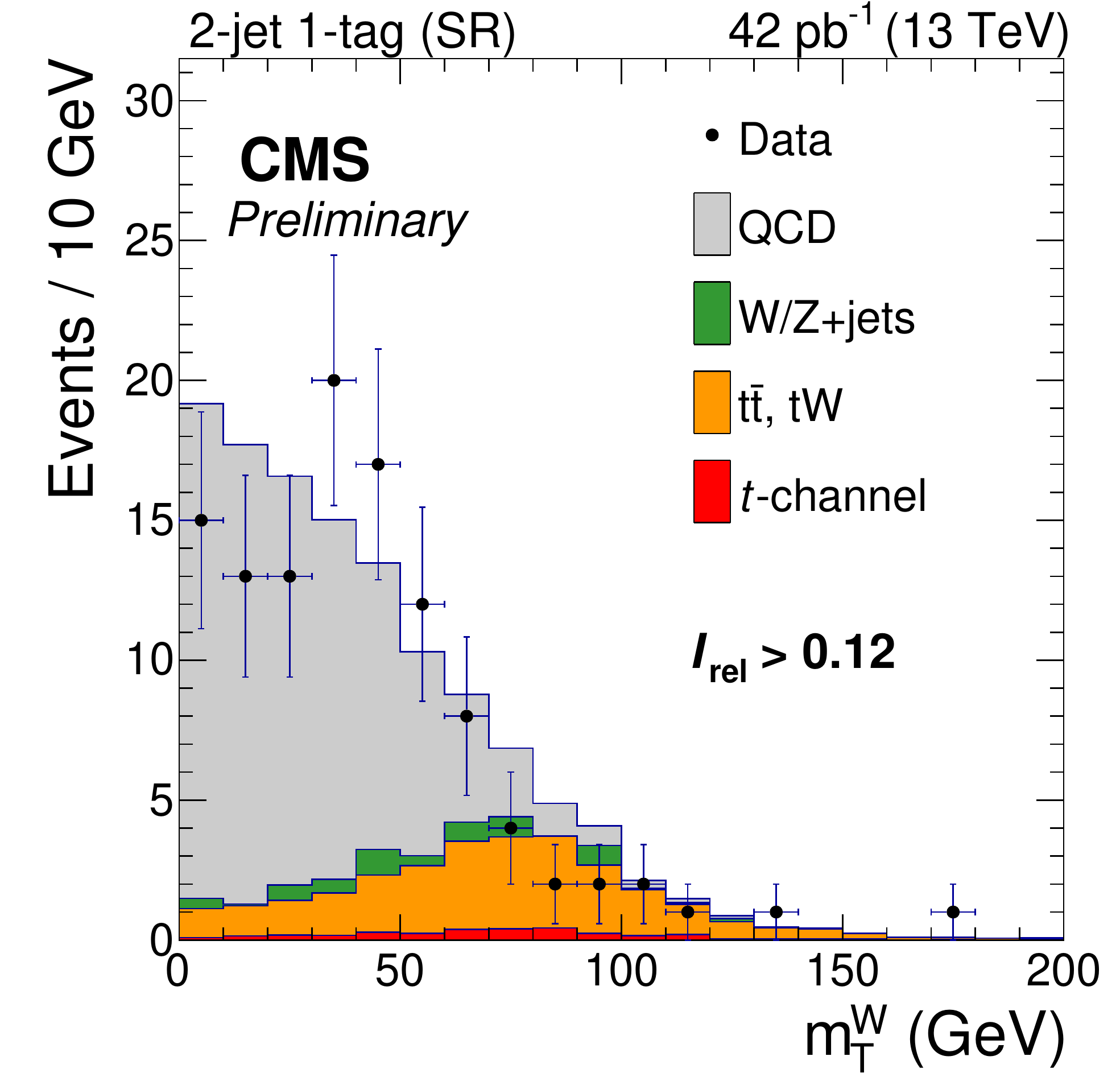}}
\end{minipage}\par\medskip
\begin{minipage}{.5\linewidth}
\centering
\subfloat[]{\label{main:a}\includegraphics[width=6cm,height=4.5cm]{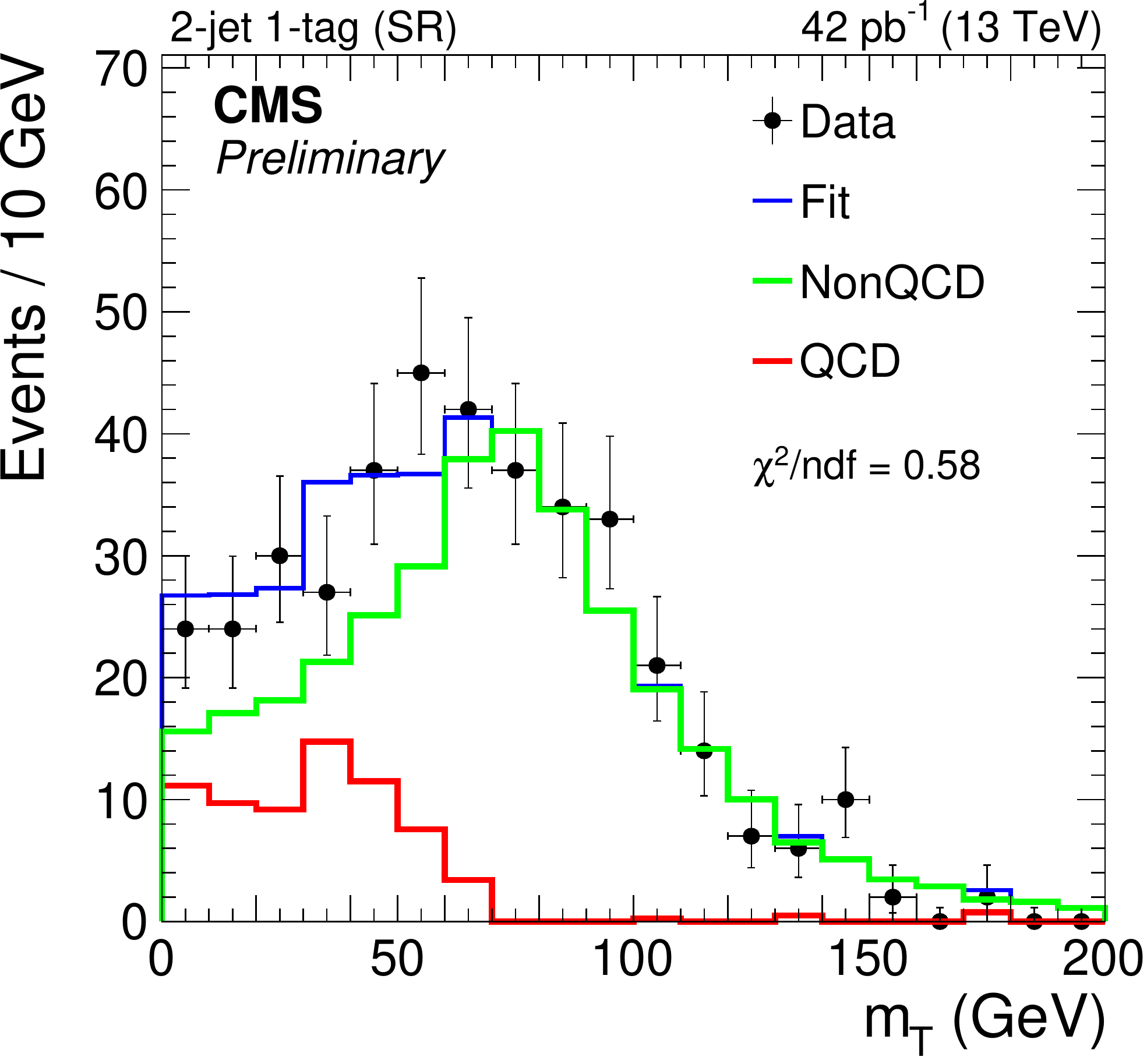}}
\end{minipage}%
\begin{minipage}{.5\linewidth}
\centering
\subfloat[]{\label{main:b}\includegraphics[width=6cm,height=4.5cm]{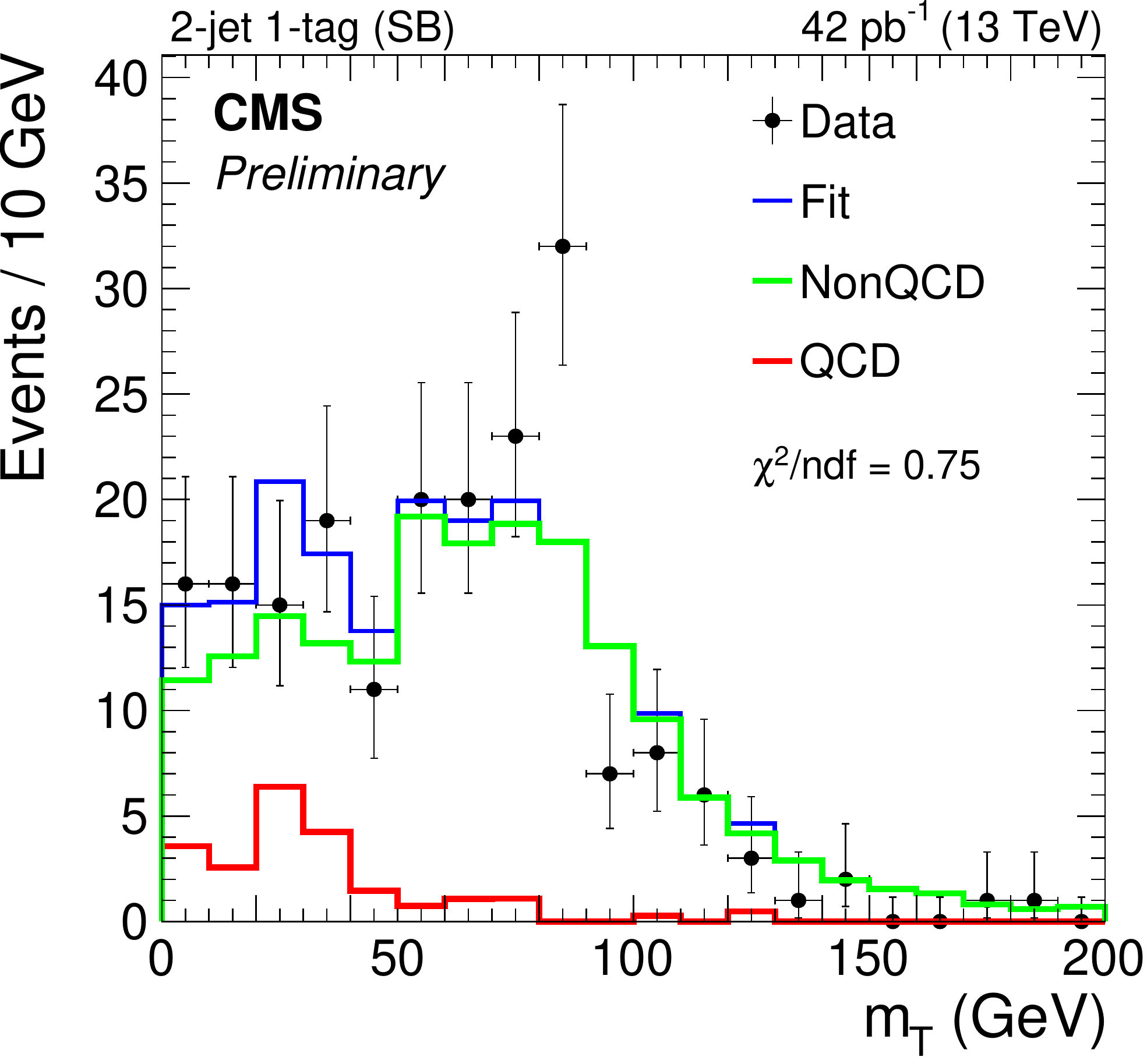}}
\end{minipage}\par\medskip

\caption{QCD estimation in 2-jet 1-tag region.}
\label{fig:main}
\end{figure}
As important as rejecting QCD away is, constraining the amount of its residual contribution to
          the SR window is equally crucial. Indeed, the reconstructed  $\rm{m_T^W}$ (Fig. \ref{fig:main}, a) shows reasonable agreement
          between data and simulation, after both QCD and non-QCD components have been scaled to the
          estimated contribution, the latter determined by the fit process. The $I_{\rm{rel}}>0.12$ separation effectively 
          results in a high statistics model for the QCD background, while non-QCD contamination is found less dominant (Fig. \ref{fig:main}, b). To exploit the discrimination power of $\rm{m_T^W}$ as much as possible, the fit (Eq. \ref{eq:fit}) is performed
          without imposing the $50\ GeV$ requirement. Thus the contribution from QCD multijet events is determined, 
          separately in SR (Fig. \ref{fig:main}, c) and SB (Fig. \ref{fig:main}, d) windows, and the QCD yield is finally evaluated
          for the $\rm{m_T^W}>50$ $GeV$ signal region \cite{cite_key3}.

\end{document}